\documentclass[a4paper,12pt]{article}
\usepackage{theorem,amsmath,enumerate}
\usepackage{latexsym}
\usepackage{amssymb,amsbsy}
\usepackage{graphicx}
\usepackage{color}
\usepackage{subfig, natbib}
\makeatletter 
\renewcommand\@biblabel[1]{#1.} 
\makeatother %
\usepackage{epstopdf}
\usepackage{psfrag}
\usepackage[pdf]{pstricks}

\usepackage{psfrag}
\usepackage[pdf]{pstricks}
\newcommand{\subfigure}{\subfloat}

\newcommand{\E}{\mbox{E}}
\newcommand{\ve}[1]{\mbox{\boldmath ${#1}$}}
\newcommand{\vesub}[2]{\mbox{{\boldmath ${#1}$}$_{#2}$}}

\newcommand{\vess}[3]{\mbox{{\boldmath ${#1}$}$_{#2}^{#3}$}}

\newcommand{\CF}{\mathcal F}

\newcommand{\eql}[2]{\begin{equation} #1 \label {#2} \end{equation}}

\newcommand{\eqn}[1]{\begin{eqnarray} #1  \end{eqnarray}}
 
\newcommand{\eqnn}[1]{\begin{eqnarray*} #1  \end{eqnarray*}}

\newcommand\expit{\mbox{expit}} 
\newcommand\logit{\mbox{logit}} 
\newtheorem{lem}{Lemma}

\newcommand{\Hline}{\hline \hline}
\newcommand{\uf}[1]{{\noindent {\bf #1}}}

\title{Simulation-based Sensitivity Analysis for Non-ignorable Missing Data} 

\author{Peng Yin and Jian Qing Shi\thanks{School of Mathematics \& Statistics, Newcastle University, Newcastle NE1 7RU, UK. Email: j.q.shi@ncl.ac.uk}\\
 School of Mathematics \& Statistics, Newcastle University }

\begin{document}

\maketitle

\begin{abstract} 
Sensitivity analysis is popular in dealing with missing data problems particularly for non-ignorable missingness. It analyses how sensitively the conclusions may depend on assumptions about missing data e.g. missing data mechanism (MDM). We called models under certain assumptions  sensitivity models. To make sensitivity analysis useful in practice we need to define some simple and interpretable statistical quantities to assess the sensitivity models.  However, the assessment is difficult when the missing data mechanism is missing not at random (MNAR).  We propose a novel approach in this paper on attempting to investigate those assumptions based on the nearest-neighbour (KNN) distances of simulated datasets from various MNAR models. The method is generic and it has been applied successfully to several specific models in this paper including meta-analysis model with publication bias, analysis of incomplete longitudinal data  and regression analysis with non-ignorable missing covariates. 
\end{abstract}


\section{Introduction}
Model uncertainty and incomplete data are common issues in statistical analysis. An assessment of uncertainty due to incomplete data or model misspecification has attracted many researchers' attention for several decades \citep[see e.g][]{CHHLSW59, VDEV69, Draper95, CL97}. Let $\theta$ be the parameter of interest and $D$ a set of observations. Conventional inference employs a model $f(D;  \theta)$ and this usually provides a consistent estimate for $\theta$. However, if part of the data are unobserved, inference based on the observed data may result in bias. In this case bias analysis can be carried out by using a so-called bias model  $f(D; \theta, \eta)$ associated with different missing data mechanisms (MDM) \citep[see e.g.][]{Greenland05}, where  $\eta$ is called the bias parameter. It is   usually difficult to estimate $\eta$ due to the lack of data or prior knowledge. For example, if the missingness is under so-called non-ignorable (or missing not at random) MDM  \citep{LR02} and $\eta$ is the parameter involved in a MDM model, then $\eta$ cannot  be estimated as it depends on the unobserved data.  In some cases it is possible to resort to follow-up studies in order to estimate $\eta$ \citep[see e.g.][]{KY11};  in many instances, however, such investigations are inherently difficult to conduct due to the nature of the study as it occurs, for example, in epidemiological studies.  Also extra bias may exist due to the lack of randomization and the independence problem between former observations and follow-up samples. 

Sensitivity analysis is one of the commonly used approaches for assessing uncertainty via a \emph{bias parameter} $\eta$ or the related \emph{bias model}; hence $\eta$ is also known as the \emph{sensitivity parameter} and its associated model the \emph{sensitivity model}. As such, we will refer to $\eta$ the sensitivity parameter in the remainder of the paper. The use of these models is prolific in many and varied areas, for example, \cite{Copas99} and \cite{CS00,Copas01a}  to explore  publication bias in meta-analysis using the Heckman model \citep{Hechman79},  \citet{MKG01} to investigate  incomplete contingency tables  and \citet{CE01, CE05} to look at local sensitivity analysis. Those discussions characterize the sensitivity analysis in different ways, but their aims are essentially the same: to examine the influence of an individual point on model-based inference. A different approach considers all possible sources of uncertainty by defining a prior density coupled with a Monte Carlo sensitivity analysis to sample `{\it bias parameters and then inverts the bias model to provide a distribution of bias-corrected estimates}' \citep[][p.269]{Greenland05}. However it is usually difficult to choose and justify a prior density.  

\cite{Copas13} pointed out that `{\it a sensitivity analysis is essentially a warning of how sensitively the conclusions of a meta-analysis may depend on key assumptions about the study selection process}'. Using the idea in a general sensitivity model $f(D;\theta, \eta)$, we need to analyse how the conclusions change when $\eta$ changes. It is usually more useful for practitioner if we can find some simple and interpretable statistical quantities to assess the sensitivity models instead of using $\eta$ directly. For example \cite{Copas01a} used the P-value for the goodness of fit to the funnel plot and the estimated number of unpublished studies. In spite of being very useful, those methods are limited to special cases and it is not straightforward to extend the idea to other problems. In this paper we propose a generic method  based on the nearest-neighbour distance between the observed data and the data simulated from  different sensitivity models. We attempt to achieve two aims: (i) exclude the models which are `far away' from the true model. We do so by comparing the KNN distances with a critical value based on a permutation test; (ii) in the set of plausible sensitivity models resulting from (i), find the most plausible model or a set of the most plausible models in terms of  minimal distance. This will result in a set of most plausible estimates in a small range for the parameter of interest $\theta$, which would provide useful information to help practitioners to  draw conclusions.  We will refer to  this method as \emph{simulation-based sensitivity analysis} (SSA).      

The remainder of the paper is organized as follows. Section 2 will first describe the idea of simulation-based sensitivity analysis, followed by a detailed discussion on how to evaluate sensitivity models using the nearest-neighbour distance and a permutation test. The method is applied to a missing data problem in meta-analysis and a longitudinal study in Section 3; both examples are illustrated with numerical results from real data problems.   Finally, a discussion is given in Section 4 with further examples presented in Appendix. 

\section{Simulation-based sensitivity analysis}
 
\subsection{Sensitivity models and sensitivity parameters} \label{sec2.1}

 Let $\CF$ be the population of complete data from which we wish to infer the parameter of interest $\theta$ using model $f(\CF; \theta)$. If a sample $D$ is drawn randomly from $\CF$,  $\hat \theta$ calculated from the model $f(D; \theta)$ is usually unbiased. However, observed data, denoted by $D_{obs}$, are often not a representative of the complete data. Conventional inference usually employs a model $f(D_{obs};\theta)$ under assumptions of missing at random (MAR) and this may result in bias since those assumptions are often invalid under some `imperfect' situations such as publication bias in meta-analysis,  measurement error with non-ignorable missingness or the use of a misspecified MDM model. The effect of bias source may be modeled with a bias model parametrized with a bias parameter. 

 Let $D=(D_{obs}, D_{mis})$ be a set of complete data including both observed and unobserved data and $R$ a missingness indicator vector which takes 1 if data is observed or 0 otherwise. The complete data model can be factorized into an extrapolation model and an observed data model as follows. 
\eql{ f(D, R| \Theta) = f(D_{mis} |D_{obs}, R, \Theta_m) f(D_{obs}, R|\Theta_o).} {eq: obs variable}
Here, $\Theta_m$ and $\Theta_o$ denote parameters indexing the models for missing data  and observed data respectively. The item $f(D_{mis} |D_{obs}, R, \Theta_m)$  is called extrapolation model since the missing data are often outside the range of the observed data. Usually the parameter of interest $\theta$ is a subset of $\Theta_0$ and the remaining components are the nuisance parameters. To simplify notations we will not distinguish $\Theta_0$ from $\theta$ and will always use the latter.  The extrapolation model cannot usually be identified unless extra assumptions are made.  The \emph{bias} or \emph{sensitivity parameters} \citep{DH08, Greenland05}, denoted by $\eta$, are a function of the parameter $\Theta_m$. These parameters are usually inestimable. But when their values are given, the full data model $f(D,R|\theta, \vesub{\Theta}{m})$ in \eqref{eq: obs variable} is identified. In another words, the estimate of $\theta$ depends on the value of $\eta$, and it may be biased if the value of $\eta$ is given wrongly. We assume one value exists,  denoted by $\eta_0$, for which  the model $f(D,R|\theta, \vesub{\Theta}{m})$ would provide an unbiased estimate of $\theta$ given $D_{obs}$ and $\eta_0$. 

\begin{figure}[!ht] 
	\begin{center}
	\includegraphics[scale=0.4, angle=-90]{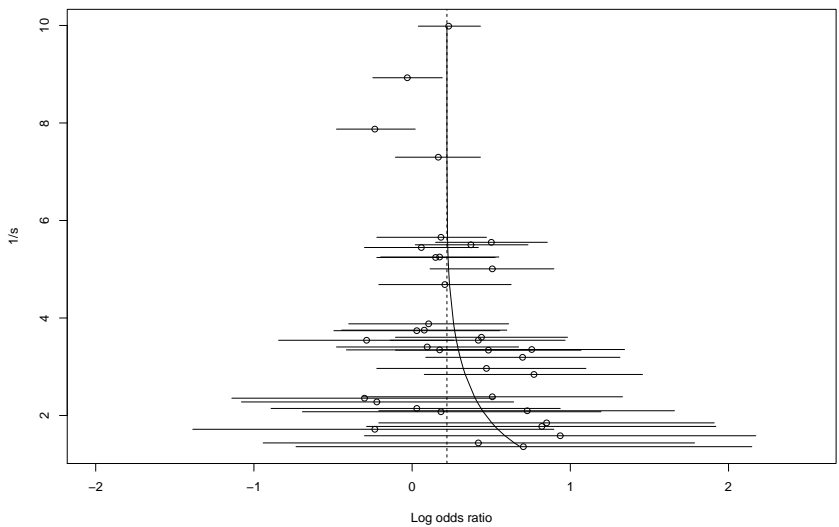}	
		\caption{Passive smoking and lung cancer data: log odds ratios and their 95\% confidence interval versus the inverse of the sample standard errors for 37 studies; the dashed line represents $\hat{\mu}=0.22$, the estimate without assuming selectivity; the solid line represents fitted values obtained from the most plausible model with $(a,b,\hat{\mu})=(-2.6,0.8, 0.165)$. }
\label{fig: lungcancer logoddsratio}
\end{center}
\end{figure}

We now illustrate the meaning of sensitivity model and sensitivity parameters using an example which assesses  the relationship between passive smoking and lung cancer   \citep{Hackshaw97}. A total of  37 published epidemiological studies are selected to investigate  the risk of lung cancer in female non-smokers whose spouses/partners did or did not smoke. 
Suppose $y_i$ is the estimated  log odds ratio reported in the $i$th study and $s_i$ is the corresponding standard error for $i=1,\ldots, n$. A random-effects meta-analysis model  is given by
\eql{ y_i = \mu_i + \sigma_i \epsilon_i, \, \mu_i \sim N(\mu, \tau^2), \, \epsilon_i \sim N(0,1), \, i =1,\ldots,n. }{eq: log odds model}
Here $\mu$ is the overall mean effect which is the parameter of interest, $\tau^2$ is  the variance measuring heterogeneity while $\sigma_i^2$ is the within-study variance and is usually replaced by the sample variance $s_i^2$. 
Figure \ref{fig: lungcancer logoddsratio} 
 presents the funnel plot of the log odds ratios showing the sign of publication bias, i.e. smaller studies give more positive results than the larger studies. In other words smaller studies with inconclusive results are less likely to be selected; see the detailed discussion in \cite{CS00}. 
 
Without assuming publication bias, the maximum likelihood estimate of the overall mean is $\hat{\mu}=0.22$. The relative risk is therefore 1.246 or the excess risk is $24.6\%$, which implies that people who live with smokers have a  $24.6\%$ excess risk to have lung cancer when compared  with those living with non-smokers. This is however an over-estimate due to publication bias as argued by \cite{CS00}. 
To address  the problem they  used the following selection model  
\eql{z_i= a+b/s_i +\delta_i, \, \delta_i \sim N(0,1), \, \mbox{corr}(\epsilon_i,\delta_i) = \rho.}{eq: pb model}
A study is selected  when $z_i >0$ (i.e. $R_i=1$). In this model  $\eta=(a,b)$ are the sensitivity parameters.  If $\eta=(a,b)$ is given, $P(z_i>0|y_i)$ can be calculated from the selection model whereas the parameter of interest $\mu$ and the nuisance parameters $(\tau^2, \rho)$ can be estimated from  the meta-analysis model \eqref{eq: log odds model} and the selection model \eqref{eq: pb model} \citep{CS00} by maximizing the following log-likelihood. 
\begin{eqnarray}
L(\mu, \tau^2, \rho) & = & \sum_{i=1}^n [ \log f(y_i|z_i>0) ] \nonumber \\
& = & \sum_{i=1}^n [ \log f(y_i) + \log P(z_i>0|y_i) - \log P(z_i>0) ]. \label{metalik}
\end{eqnarray}
However, since the number of unselected studies is unknown, $\eta=(a,b)$ is inestimable. In this example, uncertainty exists for the sensitivity parameters $(a,b)$ as well as the assumed selection model  \eqref{eq: pb model}. A sensitivity analysis investigates how sensitively the estimate of $\theta$ depends both on the choice of the selection model and the choice of the sensitivity parameter $\eta$. 

The sensitivity parameter $\eta$ in this example has no clear physical meaning. It is difficult to draw any meaningful conclusion if it is used directly.
To make sensitivity analysis useful we need to develop some simple and interpretable quantities. \cite{CS00} used the p-value of the goodness-of-fit to funnel plots   and \cite{Copas13} used the overall selection probability. Those quantities have clear physical interpretation  from which meaningful conclusions can be drawn; this idea, however, is not easily expandable to other missing data problems. In this paper, we propose a more generic method which works by comparing the observations with datasets simulated from  candidates of the sensitivity model $f(D; \theta, \eta)$. The nearest-neighbour distance is used to evaluate those models. In principle, this method can be applied to any missing data problem; we provide further details about the procedure in the next two subsections.

\subsection{Evaluating sensitivity models} \label{sec: bms}

As we have discussed previously,  $f(D,R; \theta, \eta)$ is the sensitivity model, where $D=(D_{obs},D_{mis})$, $\theta$ is the parameter of interest and $\eta$ is the sensitivity parameter. There is uncertainty for  $\eta$, but $\theta$ can be estimated when $\eta$ is given.  For example, we may estimate $\theta$ through a marginal density 
\eqn{ f(D_{obs},R|\theta, \eta) = \int f( D_{obs}, D_{mis},R|\theta,\eta) d D_{mis}. \label{profile likelihood}
}
The  meta-analysis example discussed in the previous subsection follows this approach, and $\theta$ is estimated from the marginal likelihood \eqref{metalik} for the observed data when $\eta=(a,b)$ is given.


Let $\eta_0$ be the true value.  The observed data $D_{obs}$ are therefore generated from either the full model $f(D,R; \theta, \eta_0)$ or its corresponding marginal model $f(D_{obs},R;\theta,\eta_0)$.  Hence, if the value $\eta_0$ is known, an unbiased estimate for $\theta$ can be determined  from the model and the observed data $D_{obs}$; see the discussion on the related topics in \cite{lu04}. Unfortunately  $\eta_{0}$ is usually unknown and it cannot be estimated from $D_{obs}$ under non-ignorable missingness.
Our idea considers a set of sensitivity parameters denoted by $\Gamma$ and investigate each $\eta$ in $\Gamma$ by comparing $f(D, R; \theta, \eta)$   with $f(D,R;\theta,\eta_0)$. If they are very close to each other, the candidate model would result in a (nearly) unbiased estimate under certain regularity conditions. Since  $\eta_0$ is unknown, the comparison cannot be made directly; an indirect approach simulates datasets from the candidate model and proceeds further by comparing the simulated and the observed data  (bearing in mind that the observed data $D_{obs}$ come from $f(D,R;\theta,\eta_0)$).
 

The steps of this simulation-based sensitivity analysis (SSA) procedure are as follows. 
\begin{itemize}
\item[(i)] Select one $\eta$ in $\Gamma$ or generate it from a prior distribution $p(\eta)$ if we have some prior knowledge about $\eta$;
\item[(ii)] Estimate $\hat{\theta}_{\eta}$ using the sensitivity  model $f(D,R; \theta, \eta)$ given the selected  $\eta$;
\item[(iii)] Simulate an incomplete sample $D_{\eta}^{obs}$ from the model $f(D,R; \hat{\theta}_{\eta}, \eta)$; 
\item[(iv)] Calculate distance $s(D_{\eta}^{obs},D_{obs})$; 
\item [(v)] Repeat Steps (i) to (iv) for each candidate $\eta$ in $\Gamma$. 
\end{itemize}

We have demonstrated how to calculate $\hat{\theta}_{\eta}$  from $f(D,R; \theta, \eta)$ given $\eta$ in Step (ii) using the  meta-analysis and publication bias example in the previous subsection. More examples will be discussed later in this paper. 


 In Step (iii) we first sample a complete dataset $D_\eta$  from $f(D,R; \hat{\theta}_{\eta},\eta)$ and then  remove some elements from $D_{\eta}$ in order to generate the incomplete observations $D^{obs}_\eta$; this latter dataset is comparable to the actual observations, $D_{obs}$. This process requires a missing data mechanism  to be specified by $\eta$. As there is no an unified approach  to simulate $D_\eta$ or $D_\eta^{obs}$, we discuss some viable techniques in Section \ref{sec:examples} and Appendix A.3.


Step (iv) calculates the distance between the simulated dataset $D_{\eta}^{obs}$  and the actual observations $D_{obs}$. The key here is defining  a proper  measure  in order  to measure the `closeness' or `similarity' between  the sets. This is  particularly important  for the large-dimensional case. To measure the similarity or dissimilarity between two clusters various statistical distances have been investigated in the literature. A common approach calculates the distance between each pair of data points in $D_{\eta}^{obs}$ and $D_{obs}$, and then use either the minimum  (single linkage by \citet{Sneath57}), the maximum (complete linkage by \citet{Sorensen48}) or the  average distance \citep{SM58}. In our experience, the first two distances do not work very well in SSA  while the average distance works well for some examples but the performance is not  consistent. As an alternative,  we use the $K$ nearest-neighbour (KNN) method, first introduced by \cite{FH51} as a nonparametric measure; further details are given in Appendix A.1.  This measure works well in most of the cases we have tried. 

We will discuss how to use the distance $s(D_{\eta}^{obs},D_{obs})$ to conduct a sensitivity analysis in the next subsection.  

\noindent {\bf Remark 1}. The sensitivity model, either $f(D,R| \theta, \eta)$ or $f(D_{obs},R| \theta, \eta)$, depends on the sensitivity parameter $\eta$ as well as the hierarchical structure (see equation \eqref{eq: obs variable}); so does $D_{\eta}^{obs}$. Hence, the distance $s(D_{\eta}^{obs},D_{obs})$ can be used to investigate both the plausibility of $\eta$  and the model structure. This method  is therefore appropriate for the study of misspecified models  and other related problems.

\noindent {\bf Remark 2}. Numerical problems may arise in step (iv) when calculating the distance between $D_{obs}$ and one set of $D_{\eta}^{obs}$. A strategy that works in those instances makes use of an average distance by sampling more than one set from  $D_{\eta}^{obs}$ for a given $\eta$. We will be referring to the number of replicates taken to build the average distance as a Monte Carlo sample size or simply MC size.

\noindent {\bf Remark 3}. KNN distance is just a way to measure similarity between $D_{obs}$ and  $D_{\eta}^{obs}$. It performs well for the examples discussed in this paper. However, some other distances or measures may be used for different types of data or problems. 

\subsection{Sensitivity analysis}


The distance $s(D_{obs}, D_{\eta}^{obs})$ measures the `closeness' between the sensitivity model and the `true' model. Those models giving rise to large values of this metric can simply be discarded as they are very unlikely. In order to define a test criterion more formally, we borrow the idea of an 
`achieved significance level' (ASL) introduced by \citep{Fisher71} using permutation tests. Let now $D_1$ and $D_2$ denote a pair of permutation samples drawn from the combined dataset $D^*=(D_{obs}, D_\eta^{obs})$ whose distance is $s^*=s(D_1, D_2)$. We define the ASL for SSA as:
\begin{equation} \label{eq: asl}
ASL_{\eta}= \mbox{Pr}_{H_0} \{ s^* \ge s(D_{obs}, D_{\eta}^{obs}) \}, 
\end{equation}
where $H_0$ is the null hypothesis stating that $D_{obs}$ and $D_{\eta}^{obs}$  come from the same distribution. When we need to resort to average distances  to tackle numerical instabilities as per Remark 2, permutation samples are generated over the multiple  $D_{\eta}^{obs}$ with a given MC size. The ASL is the proportion of the pairs of the permutation samples for which the distances are larger than the distance between  $D_{obs}$ and $ D_\eta^{obs}$. Thus ASL works similarly to a p-value with larger values taken as being in favour of $H_0$. 

Other approach to work out the ASL  uses the concept of `internal distance'; in that case, the  permutation sample $(D_1,D_2)$ is drawn from the $D_\eta^{obs}$ instead of the combined dataset $D^*$.   

For a given significance level (e.g. $\alpha=5\%$), a plausible set of sensitivity parameters is defined as
$ \Gamma_\alpha = \{ \eta: ASL_{\eta} > \alpha \}. $
Models not included in $\Gamma_\alpha$ should be avoided. 

For all the models within $\Gamma_\alpha$, the distance calculated for each $\eta$ can be used as a further guide in deriving an estimate for $\theta$.  For instance, we define the `most plausible value' of $\eta$ that  gives rise to the smallest distance and term the corresponding estimate of $\theta$ as the `most plausible estimate'. It is important to emphasize, however, that this estimate may not be consistent regardless of the sample size. Hence,  caution should be exercised even when the most plausible model has been chosen; we provide further discussion about this topic in the examples.

\section{Examples} \label{sec:examples}

We will discuss two missing data problems in this section, including publication bias problem in meta-analysis and incomplete longitudinal data.   
 Techniques on how to generate $D_{\eta}$ from $f(D; \theta, \eta)$ and how to sample $D_\eta^{obs}$ from $D_\eta$ will also be discussed. More examples will be given in Appendix, including mean estimation with non-ignorable missing data and non-ignorable missing covariates.

\subsection{Publication bias in meta-analysis}

We carry out the meta-analysis for the example of passive smoking and lung cancer  discussed in Section \ref{sec2.1}.  In this example, $\mu$ is the parameter of interest, $(\tau^2, \rho)$ are the nuisance parameters and $\eta=(a,b)$ are sensitivity parameters which are inestimable. When $\eta=(a,b)$ are known, the marginal likelihood for the observed data is given by  \eqref{metalik}, i.e.
\eqnn{  L(\mu,\tau,\rho| a,b) 
	&=&\sum_{i=1}^n {[\log f(y_i|z_i>0, s_i)]} \\
	&=& \sum_{i=1}^n{\left[-\frac{1}{2} \log (\tau^2+\sigma_i^2) - \frac{(y_i-\mu)^2}{2(\tau^2+\sigma_i^2)} -\log \Phi(a+\frac{b}{s_i}) +\log \Phi(v_i) \right]}
	}
	where \[v_i=\frac{a+\frac{b}{s_i}+\tilde{\rho}_i \frac{y_i-\mu}{(\tau^2+\sigma_i^2)^{1/2}}}{(1-\tilde{\rho}_i^2)^{1/2}} , \, \tilde{\rho}_i = \frac{\sigma_i}{(\tau^2+\sigma_i^2)^{1/2}} \rho, \]
	and $\Phi(\cdot)$ is the distribution function of the standard normal distribution. Parameters $(\mu, \tau^2, \rho)$ can be estimated by maximizing the above marginal likelihood given the observed data. 
	
\begin{figure}[!ht] 
	\begin{center}
\subfigure[$\hat{\mu}$]{\includegraphics[scale=0.3,angle=-90]{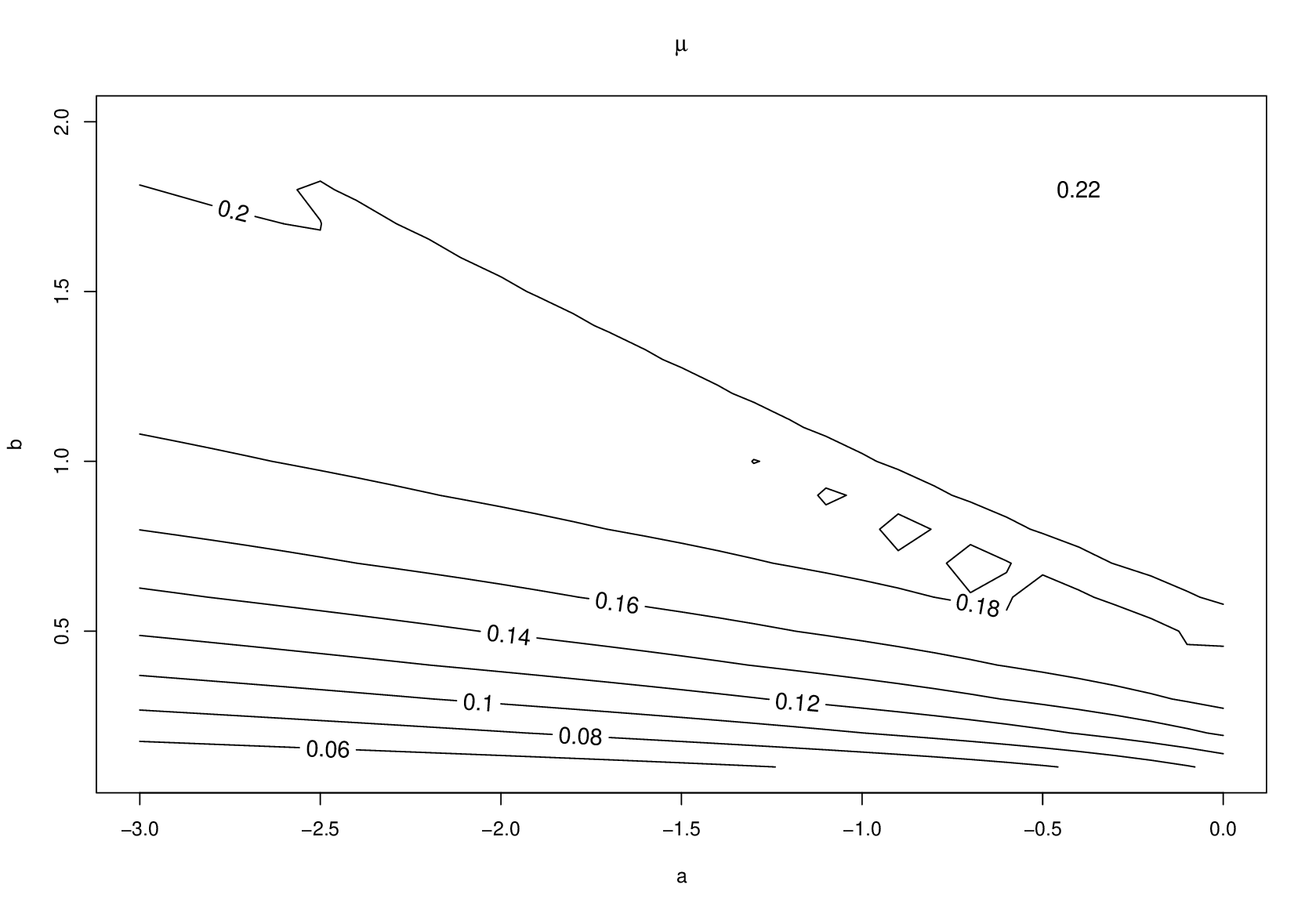}  } 
\subfigure[average distance]{\includegraphics[scale=0.3,angle=-90]{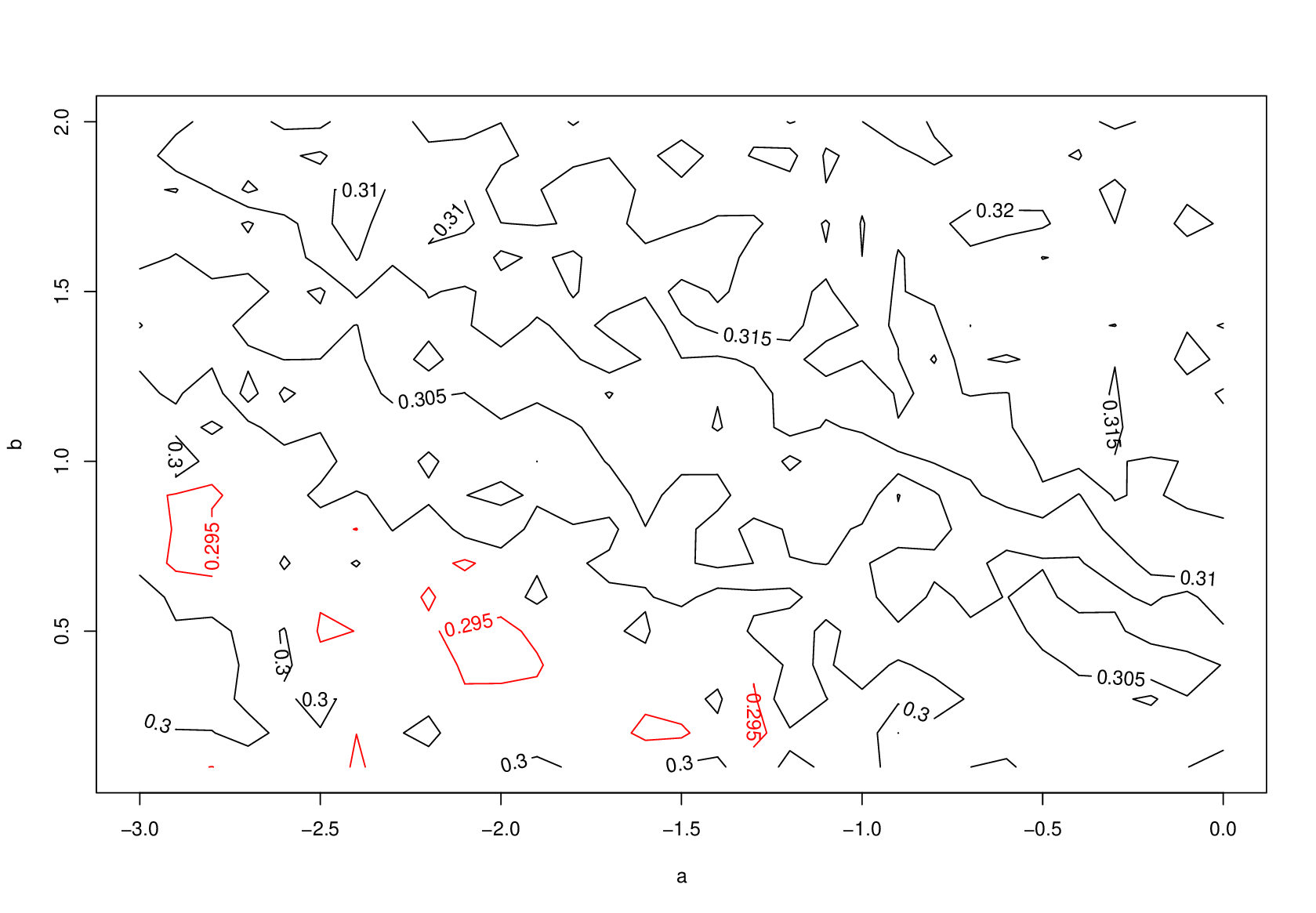} \label{fig: lungcancer BMS}} %

	\caption{Contours plots for passive smoking and lung cancer data. In (b), the area in red  corresponds to the most plausible values for $\eta$ with the average distances being all very close to the smallest distance of 0.295.}
		\label{fig: lungcancer SA}
\end{center}
\end{figure}

	A sensitivity analysis is conducted to investigate how the estimates of $\mu$ change for different values of $\eta=(a,b)$. The selection of possible values of $a$ and $b$ is based on a range of selection probabilities for the largest  and the smallest studies as suggested in \cite{Copas01a}. This region is $a \in (-3,0)$ and $b \in (0.1,2)$  as shown in Figure \ref{fig: lungcancer SA} (a),  from which a grid of values can be chosen accordingly to conduct the sensitivity  analysis. The area towards  the bottom left indicates a high potential for publication bias; this reduces as we move towards the top right of the figure.  For each selected pair of ($a,b$), $\mu$ is estimated using the  maximum likelihood method. A contour plot is presented in Figure \ref{fig: lungcancer SA} (a). At the top right $\hat{\mu} =0.22$ which agrees with the estimate from a model without assuming publication bias but this value falls as the sensitivity parameters move towards the bottom left corner. A question which arises naturally is: which estimate should we use? In this regard, \cite{CS00} and \cite{Copas01a} have suggested several statistical quantities which are interpretable and from which some answers can be found to the aforementioned question. 
	
	We now use the SSA method to conduct a sensitivity analysis.   For each pair of  $(a,b)$ we calculate the estimates $(\hat{\mu}, \hat{\tau}^2, \hat{\rho})$ and simulate a complete dataset using meta-analysis model \eqref{eq: log odds model}; denoted the simulated data by $D=\{(y_j, s_j), j=1, \ldots, N\}$. Given the known values of $(a,b)$ and the estimates of the parameters, we subsequently  draw  random numbers of $z$  from the selection model \eqref{eq: pb model}. Note that $z$ is correlated with $y$ and the study is not selected if $z<0$. The selected studies form the observed data  $D_{(a,b)}^{obs}$. This dataset is compared with the actual dataset $D_{obs}$ by using KNN distance $s( D_{(a,b)}^{obs}, D_{obs})$. In accordance with Remark 2, Section 2.2, we have used the average distance with 1000 replicates for each pair of $(a,b)$ in order to remove any numerical instability (i.e. the MC size is 1000).  The contours of the average distances are shown in  Figure \ref{fig: lungcancer SA} (b).

Using the permutation method, discussed in Section 3.3, we can calculate the ASL for each value of $\eta$ and discard all pairs of sensitivity parameters whose ASL values are smaller than 0.05. The remaining pairs form a plausible set of sensitivity parameters which results in estimates for $\hat{\mu}$ ranging from 0.043 to 0.197. The upper bound for this estimate indicates an excess risk of 21.7\%. The conventional method (without assuming publication bias) results in an excess risk of 24.6\%, thus overestimating the SSA upper bound by 13\%.   

	
 The most plausible value (smallest distance) is achieved when $(a,b) =(-2.6, 0.8)$ which leads to  the estimate of $\hat{\mu}=0.165$ or an  excess risk of 17.9\%. The fitted line is shown in Figure 1, showing a much better fit than the dashed line which is obtained from the model without considering publication bias.

As we have mentioned before hand, caution is needed when working with the most plausible value. This can be seen by looking at Figure 2(b), where the average KNN distances are all very close to the shortest distance of 0.295 for several values of the sensitivity parameters. The estimates of $\mu$ for those sets range between 0.086 and 0.183 which lead to excess risks in the interval between 9\% and 20\%.


Based on the 37 studies collected in the meta-analysis and the simulation-based sensitivity analysis, we can conclude that the excess risk is not likely to be higher than 21.7\% or smaller than 4.4\%. Most likely it will take values between   9\% and 20\%. This result is consistent with the findings in \cite{CS00}. The latter is based on goodness-of-fit test  but the method is not easily extendable to other missing data problems. The SSA method proposed in this paper is more general and can be used to deal with different types of problems.


\subsection{Incomplete Longitudinal Example}
We next consider a follow-up randomized study with $M$ scheduled repeated measures. Let $Y_m$ be the outcome measured at visit $m$; we use the notation $\vess{Y}{m}{-} = (Y_1, \ldots, Y_m)$ to denote the history of the outcomes up to visit $m$ and $\vess{Y}{m}{+} = (Y_{m+1}, \ldots, Y_M)$ as the future outcomes after visit $m$. Under monotone missingness, if the outcome at visit $m$ is missing then all outcome after that visit are all missing. We denote $R_k$ as the indicator that $Y_k$ is observed, which equals to 1 if $Y_k$ is observed and 0 otherwise. We assume  that $Y_1$ is observed on all individuals ($R_1=1$) and let $d$ be the last visit on which a subject has a measure; it follows that the observed data for that subject is $\vess{Y}{d}{-} = (Y_1, \ldots, Y_d)$, where $d \le M$. We are interested in the inference of the mean $\mu=E(Y_M)$, the intended outcome at the final scheduled visit. 

The probability of dropping out at visit $k$ can be modeled by the following logistic model: 
\[  \logit \{ 1- P(d=k | d \ge k, \vess{Y}{k}{-}, \vess{Y}{k}{+}) \} = h(\vess{Y}{k}{-}) + \eta Y_{k+1} \]
where $h(\cdot)$ is an unknown function. Either parametric or nonparametric approaches can be used.  A special case is the following linear model: $h(\vess{Y}{k}{-}) = \eta_{0} + \eta_{1} Y_k$. 
This model describes the probability of dropping out at visit $k+1$ by a logistic linear model which depends on the latest recorded data $Y_k$ and the current unobserved data $Y_{k+1}$. If $\eta=0$ this is a MAR model. One would make this choice if there is evidence or belief that the missingness associated to $Y_{k+1}$ can be entirely encoded by the recorded history observations $\vess{Y}{k}{-}$. This is certainly a strong assumption in longitudinal study and usually very difficult to justify. We now use the SSA method proposed in this paper to investigate the model under a MNAR mechanism, and try to find a plausible value of $\eta$ in the neighbourhood of 0 (where 0 corresponds to the MAR model). 

Let us first consider a simple case. Suppose that we just have two visits, i.e. $M=2$. There is no missing data for the first visit, i.e., $Y_1$ is observed for all the subjects. But part of the data are missing in the second visit. Let $R$ be  the missing indicator for $Y_2$. A semi-parametric logistic regression model can be defined by:
\begin{equation} P(R=1|Y_1,Y_2) = \expit (h(Y_1) + \eta Y_2),
\label{longpi}
\end{equation}  
where $h(Y_1)$ is a nonparametric model and $\expit(t)=e^t/(1+e^t)$. The parameter $\eta$ is not identifiable from the observed data only since it also depends on the missing part of $Y_2$. 

We need the following lemma to apply SSA to the above problem. 

\begin{lem} \label{lem: selection model}
Let $T$ be the statistical variables which are partly observed.  
Given the observed data $D_{obs}$,   the conditional density of the missing part of $T$ (i.e. when $R=0$) can be expressed by
\eql{ f(T|D_{obs}, R=0) = f(T|D_{obs}, R=1) \frac{Q(D)}{E(Q(D)|D_{obs},R=1)},}{eq: c d selection}
where \eql{ Q(D)= \frac{P(R=0|D)}{P(R=1|D)}}{qd}
is the conditional odds of missing data.  
\end{lem}
The proof is given in Appendix A.2. 

 Using \eqref{longpi} and the above lemma, we have 
\eql{ f(Y_2|D_{obs}, R=0) = f(Y_2|D_{obs}, R=1) \frac{\exp (-\eta Y_2)}{\E(\exp (-\eta Y_2)|D_{obs},R=1)}.}{longselect}
In the above formula, the conditional density of the observed part of $Y_2$,  $f(Y_2|D_{obs}, R=1)$,  can be obtained parametrically or non-parametrically, for example, using a generalized additive model \citep{GAM} or Gaussian process regression \citep{Jian_Gaussian}. 
 Thus the only uncertainty in formula \eqref{longselect} is the sensitivity parameter $\eta$ which cannot be estimated from the observed data.    

The missing data can therefore be generated from  \eqref{longselect} for each given $\eta$. This forms a complete dataset for $Y_2$. We can further select a subset using the selection model \eqref{longpi} which we denote as  $Y_{2,\eta}^{obs}$. And, finally, this subset can be compared with the real observed data $Y_{2}^{obs}$ by using the KNN distance. Thus we can conduct a sensitivity analysis for $\eta$. The details  of this procedure are worked out in the  next subsection with a real example. 

Let us now move to discuss the general case with $M$ visits. Assume that the data have been recorded for all the subjects up to the $k$-th visit but with some missing data afterwards. Let $R_{k+1}$ be the missing data indicator for the $(k+1)$-th visit. We use a logistic regression model similar to \eqref{longpi} as the MDM model:
\begin{equation}  
P(R_{k+1}=1|\vess{Y}{k}{-},Y_{k+1}) = \expit (h(\vess{Y}{k}{-}) + \eta_{k+1} Y_{k+1}).
\label{longpi1} \end{equation}
Using Lemma 1, we have 
\begin{equation}
f(Y_{k+1}|D_{obs}, R_{k+1}=0) = f(Y_{k+1}|D_{obs}, R_{k+1}=1) \frac{\exp (-\eta_{k+1} Y_{k+1})}{\E(\exp (-\eta_{k+1} Y_{k+1})|D_{obs},R=1)}.
\label{longselect1} \end{equation}
We can generate the missing part of $Y_{k+1}$ using the above density function for any given $\eta_{k+1}$ and  then use those generated data to fill in the missing observations in $Y_{k+1}$; we denote the data up to the $(k+1)$-th visit as $\vess{Y}{k+1,\eta_{k+1}}{-}$. Note that the imputed data for missing part of $Y_{k+1}$ depends on the sensitivity parameter, so the notation is sub-indexed by $\eta_{k+1}$.  

The procedure goes on to generate the missing data for the $(k+2)$-th visit in a similar way using formulas \eqref{longpi1} and \eqref{longselect1} slightly adapted by  
\[
P(R_{k+2}=1|\vess{Y}{k+1,\eta_{k+1}}{-},Y_{k+2}) = \expit (h(\vess{Y}{k+1,\eta_{k+1}}{-}) + \eta_{k+2} Y_{k+2}),
\]
and
\[
f(Y_{k+2}|D_{obs,\eta_{k+1}}, R_{k+2}=0) = f(Y_{k+2}|D_{obs,\eta_{k+1}}, R_{k+2}=1) \frac{\exp (-\eta_{k+2} Y_{k+2})}{\E(\exp (-\eta_{k+2} Y_{k+2})|D_{obs,\eta_{k+1}},R=1)},
\]
where $D_{obs, \eta_{k+1}}$ includes all the observed data and the inputed data for missing part in the $(k+1)$-th visit. The above density function can be used to generate missing data and fill in the missing part for the $(k+2)$-th visit. Keeping up with the notation,  $\eta_{k+2}$ is the sensitivity parameter in this step.

The procedure continuous until all the missing observations are filled in. In this case, the generated data depends on a sensitivity parameter vector  $\ve \eta=(\eta_{k+1}, \ldots, \eta_M)^T$. For each visit starting from the $(k+1)$-th, we can form a subset of selected data based on the generated data and the selection model, and then compare it with the actual observations. This can be used to conduct a simulation-based sensitivity analysis for the parameter $\ve \eta$. 

\uf{Remark 4.} The aforementioned method may be very  time consuming if the dimension of $\ve \eta$ is large. In this case, we may focus on a few visits which have a large missing rate and the MDM may be non-ignorable. We can then use SSA method for those visits by considering a grid of $\ve \eta$, and assume a MAR MDM and then use multiple imputation for the other visits.         

\subsubsection{Children weight data -- a bivariate outcome with dropout}
Here we will apply the simulation-based sensitivity analysis to a real example. The 1970 British Cohort Study (BCS70) follows the lives of more than 17,000 people born in England, Scotland and Wales. Two sub-samples in the first 5 years of birth were collected: a 22-month subsample \citep{BCS70}  and a 42-month subsample \citep{BCS73}. We are interested in the average Children' weight at 42-month from birth. The 22-month subsample has a total of 2348 individuals on which information was collected; the mean and standard deviation (SD) for those observations are 11.92 and 1.56kg respectively. Likewise, the 42-month subsample has 1856 observations (i.e. data on 492 people are missing for unknown reasons) with a mean and SD of 15.04 and 1.98kg respectively.  We believe that the dropping out mechanism is associated with the first outcome ($Y_1$) and may be also associated with the current outcome ($Y_2$).

We now conduct a simulation-based sensitivity analysis as discussed in the previous subsection. We choose $\eta$ within the interval (-0.2, 0.2) and consider any  values beyond this range as  not very likely. This is because for example, if $\eta<-0.2$ , then at least 94\% of the 494 unobserved Children would have weights greater than 20 kg which is not not a realistic outcome when considering the mean of 15.04 and standard deviation of 1.98 for the observed data. 

In this example, we simply use a linear regression model 
to obtained the density function of observed $Y_2$ given the observed data, and a linear model: $h(Y_1)=\beta_0 +\beta_1 Y_1$ in \eqref{longpi}. Then, the parameters $(\beta_0, \beta_1)$ can be estimated from the observed data when the value of sensitivity parameter  $\eta$ is given. 

\begin{figure}[!ht] 
	\begin{center}
\subfigure[$\hat{\mu}$]{\includegraphics[scale=0.4]{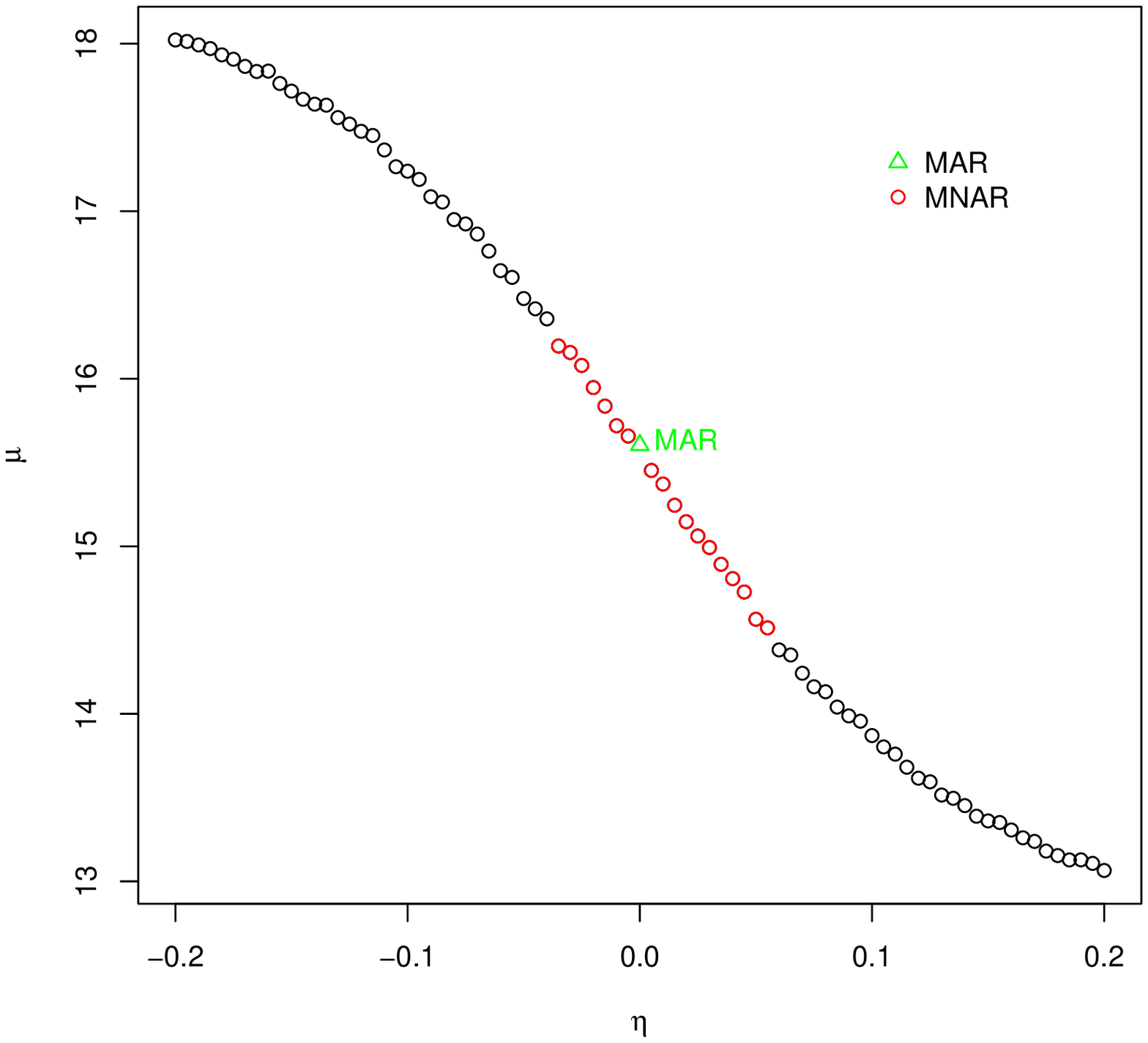} \label{fig: sa weight}  } 
\subfigure[average distance]{\includegraphics[scale=0.4]{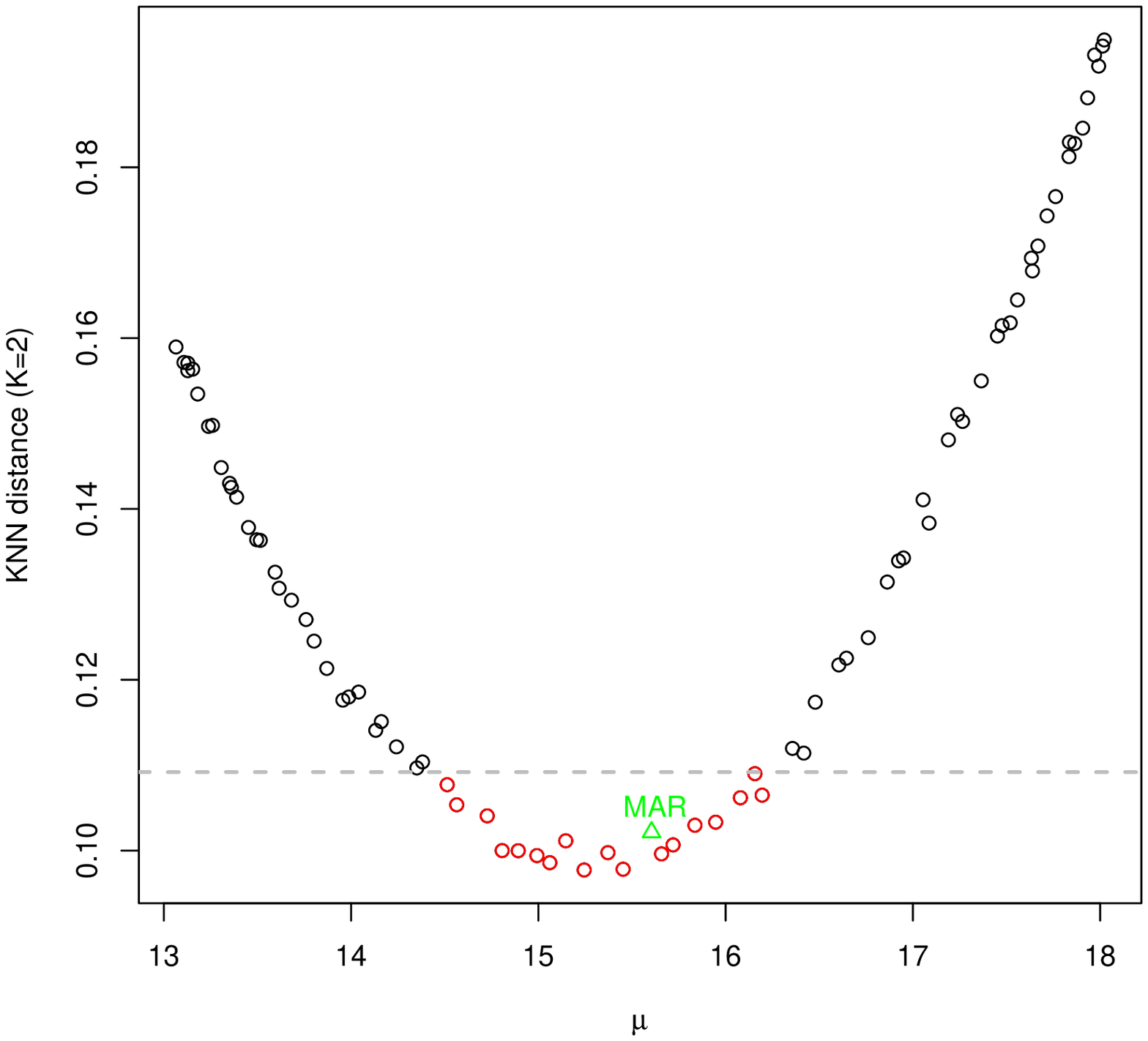} \label{fig: ssa weight}} %
	\caption{Children weight data: estimation of mean weight $\mu$ and the average KNN distance for a set of sensitivity parameter; `MAR' stands for the model with MAR assumption. }
		\label{fig: weight SA}
\end{center}
\end{figure}

The final results are shown in  Figure \ref{fig: weight SA}. The plausible set indicates that the mean estimation of the 42 month weight is in the range of (14.51, 16.19) with the most plausible value of 15.22. The average distances are presented in Figure \ref{fig: ssa weight},  providing further evidence that the sensitivity models near the bottom (corresponding to the most plausible model) are much closer to the true model than the others.  

It is interesting to note that the sensitivity model with a MAR assumption ($\eta=0$) is also included in the plausible set as shown in Figure \ref{fig: weight SA}. Although the MAR model is not the model having the shortest distance to the true model, there is no evidence to reject it. Or in other words, the non-ignorability of the missingness is not  severe. This is a by-product of using the SSA method.

\section{Discussion}

Missing data can usually be modelled by a sensitivity model $f(D,R;\theta, \eta)$, but the estimate of $\theta$, the parameter of interest, is often biased unless the true value of sensitivity parameter $\eta$ is given. Unfortunately $\eta$ is often inestimable based on the observed data  $D_{obs}$ particularly for non-ignorable missing data. A sensitivity analysis investigates how sensitively the estimate of $\theta$  depends on the choice of $\eta$. To get some meaningful results, we need to use some statistical quantities which have clear physical interpretation. The simulation-based sensitivity analysis method we proposed is based on the KNN distance. It can be applied in a general fashion and can help in gaining new and further insights into the problem at hand. The method can be applied to a great variety of missing data and model misspecification problems as we have shown in four different examples throughout this manuscript.  The numerical studies show that meaningful results can be obtained for each of those different problems.

One of the key steps when conducting a simulation-based sensitivity analysis is to be able to simulate missing  data when $\eta$ is given. Although this may be straightforward for some problems such as the meta-analysis with publication bias discussed in Section 3.1, in many other cases it will not be a trivial step; an example of this was shown in Section 3.2 which required of a somewhat elaborate process to tackle the generation of the missing observations. Other models can be treated similarly although a different array of techniques may be needed as shown in the extended examples provided in Appendix. 

Whereas the most plausible value of $\eta$ will provide a final result, in certain occasions it may be more advisable to work with a range of solutions.  As we discussed for the meta-analysis example in Section 3.1, there may be a set of $\eta$ for which the corresponding KNN distances are all very close to the shortest distance. In those cases, we recommend using a set of estimates within a small range rather  than a single value.

\section*{Acknowledgment}
The authors would like to thank the editor, associate editor and two reviewers for their critical and constructive comments and suggestions.  
The authors are grateful to the Centre for Longitudinal Studies, Institute of Education for the use of the data and to the UK Data Archive and Economic and Social Data Service for making them available; they however bear no responsibility for the analysis and interpretation of the data presented in this paper. The authors also thank Dr Javier Serradilla of Newcastle University, UK for helpful suggestions to improve the presentation of the paper.   

\bibliography{Reference}

\newpage 
\section*{Appendix}
\subsection*{A.1 K nearest-neighbour (KNN) distance} \label{knn}
Two observations $x_i$ and $x_j$ are defined to be neighbours if \citep[see Definition 1, page 364][]{WL83}:
 \[d(x_i,x_j) \le d_k(x_i) \, \mbox{or} \, d_k(x_j) ,\]
 where $d$ is the Euclidean metric and $d_k(x_i)$ is the $k$th nearest-neighbour distance to point $x_i$. Prior to defining the distance between two clusters using the KNN method, the distance between an individual point and a cluster needs to be formalized first. Given a cluster $D=\{x_i, i=1, \ldots, n \}$, an individual observation $x$ is said to be neighbour of cluster $D$ if there exists at least one point $x_i$ in cluster $D$ that
 \[d(x_i,x) \le d_k(x_i)  .\] 
 where $d$ is the Euclidean metric and $d_k(x_i)$ is the distance of $k$th nearest-neighbour within the cluster  to the point $x_i$.

Now we can define the similarity and the distance between two clusters. Let 
 $D^*=\{ x_j^*, j=1, \ldots, m\}$ be another set; we expect that most of observations in $D^*$ are in the nearest-neighbour of $D$ if the two clusters are quite similar or in 'close proximity'. A measure of how close both cluster are is given by  the proportion  of  observations in  $D^*$. This is formally defined by the  average $\E(I_1)$ where its $j$-th element has the form: 
\begin{equation} \label{eq:indicator for pnnn1}
I_1^{(j)}=\left \{ \begin{array}{cc}
1, &  \sum_{x_i \in D}\{I(d(x_i,x_j^*) < d_k(x_i)\} >0; \\
0, & \mbox{otherwise}. \end{array} \right.  \ \mbox{ for } j=1, \ldots, m.
\end{equation}
Here  $I$ is indicator function which takes 1 if the condition is satisfied and 0 otherwise, and thus $\sum_{x_i \in D}\{I(d(x_i,x_j^*) < d_k(x_i)\}$ takes an integer in the set $\{0, 1, 2, ..n \}$. Only when it equals to 0, the observation $x_j$ is not the nearest-neighbour of cluster $D$. 

 Similarly, we can define the proportion $\E(I_2)$ of the points in $D$ with
\begin{equation} \label{eq:indicator for pnnn2}
I_2^{(i)}=\left \{ \begin{array}{cc}
1, &  \sum_{x_j^*\in D^*}\{I(d(x_i,x_j^*) < d_k(x_j^*) \} >0; \\
0, & \mbox{otherwise}. \end{array} \right. \ \mbox{ for } i=1, \ldots, n.
\end{equation}

 The average of $E(I_1)$ and $E(I_2)$ \[ s(D, D^*) = \frac{1}{2} (E(I_1) + E(I_2))\] is considered as a similarity measure between $D$ and $D^*$ and $(1-s(D, D^*))$ is used as the `KNN distance'.

Other measures such as Mahalanobis metric may also be used.

\subsection*{A.2 Proof of Lemma \ref{lem: selection model}} \label{app: proof of cd}

Using Bayes Theorem, we have
\[
f(T|D_{obs}, R=0) = \frac{P(R=0|D) f(T|D_{obs})}{P(R=0|D_{obs})}.
\]
Similarly, we have
\[
f(T|D_{obs}, R=1) = \frac{P(R=1|D) f(T|D_{obs})}{P(R=1|D_{obs})}.
\]
This leads to the following equation
\[
f(T|D_{obs}, R=0) =f(T|D_{obs}, R=1) \frac{P(R=0|D)}{P(R=1|D)} \frac{P(R=1|D_{obs})}{P(R=0|D_{obs})}.
\]
Let  $Q(D)$ be the one defined in \eqref{qd}, then we have
\eqnn{
E(Q(D)|D_{obs},R=1) &=& \int \frac{P(R=0|D_{obs},T)} {P(R=1|D_{obs},T)} f(T|D_0,R=1) dT \\
		&=& \int \frac{f(T|D_0,R=1)}{P(R=1|D_{obs},T)} P(R=0|D_{obs},T) dT \\
		&=& \int \frac{f(T|D_{obs})}{P(R=1|D_{obs})} P(R=0|D_{obs},T) dT \\
		&=& \int f(T|D_{obs},R=0) \frac{P(R=0|D_{obs})}{P(R=1|D_{obs})} dT \\
		&=& \frac{P(R=0|D_{obs})} {P(R=1|D_{obs})}.
}
This proves the Lemma.

\subsection*{A.3 Additional examples} 
\subsubsection*{A.3.1 Mean estimation with non-ignorable missing data} \label{sec: mean exmple}
 Assume that a continuous variable $X$ has mean $\mu$ and variance  $\sigma^2$. The population  mean $\mu$ is of interest. The complete dataset is $D=(X_{obs}, X_{mis})$ where the MDM  depends on the missing value, and thus it is a non-ignorable missing data problem.  Let $R$  be missing data indicator and further assume that the MDM is modeled by a logistic  model
 \eql{ P(R=1|X=x) = \expit\{\eta (x + \lambda)\} }{eq: MDM model}
where $\expit (x) =\exp(x)/(1+\exp(x))$ and  $\lambda$ is assumed to be known (it can be estimated if the proportion of the missing data is known). The choice of  $\eta=0$ specifies that $R$ and $X$ are independent (i.e., MAR). 

The parameter of interest is  $\mu= E(X)$,  the mean of the complete data, can be expressed by
\[
\mu =E(X|R=1) P(R=1)+E(X|R=0) P(R=0)=\pi \mu_1 + (1-\pi) \mu_2,
\]
where $\pi=P(R=1)$ is the proportion of the observed data; $\mu_1$ and  $\mu_2$ are the means of the observed and the missing data respectively. So the evaluation of $\mu_2$ is the main task. Using Bayes theorem,  
we have \eqn{ f(x|R=0) &=& \frac{P(R=0|x) f(x)} {P(R=0)} \nonumber \\
			&=& f(x|R=1) \frac{P(R=1)}{P(R=0)} \frac{P(R=0|x)}{P(R=1|x)}. \label{eq: X|R=0}
	}
Denote that $\pi_x=P(R=0|x)$, then
\[ \frac{P(R=0|x)}{P(R=1|x)} = \frac{1-\pi_x}{\pi_x} = \frac{1}{\exp(\eta(x+\lambda))}\]
is the odds of  missing when $X=x$. The second equation comes from MDM model \eqref{eq: MDM model}. The mean of the missing data can therefore be expressed by
\eqnn{\mu_2 &=& \int x f(x|R=0) dx \\
		&=& \int x f(x|R=1) \frac{\pi}{1-\pi} \frac{1}{\exp(\eta(x+\lambda))} dx \\
		&=& \frac{\pi}{1-\pi} E_{X|R=1} \left [\frac{X}{\exp(\eta(X+\lambda))} \right] .
}
In this example, $\eta$ is the sensitivity parameter.  It is clear that we are unable to estimate this parameter from the observed data alone as it depends on the missing data as well.

Equation  \eqref{eq: X|R=0} is the key to conduct a simulation-based sensitivity analysis as it is needed in step (iii) of the SSA procedure to simulate data.

We now conduct a simulation study to demonstrate the use of the SSA approach in order to address  the non-ignorable missing data problem.  The data is generated from a normal distribution $X \sim N(\mu, \sigma^2)$ with the true values selected as $\mu=0, \sigma^2 =1$ and $\eta=-1, \lambda=0$ in model (\ref{eq: MDM model}), indicating an average missing proportion of about 50\%. Sample size of the complete data is 100. In this example $\lambda$ is assumed to be fixed and $\eta$ is treated as a sensitivity parameter. The SSA approach is designed as follows. We first select  a series of $\eta$  chosen from the interval of (-5,5).  For each selected $\eta$, we evaluate the density $f(x|R=0)$ by \eqref{eq: X|R=0} and then use the density function to sample the missing data, denoting the imputed values as $x_{mis, \eta}$. Thus, $D_{\eta}=(x_{obs}, x_{mis, \eta})$ forms a simulated complete dataset. To compare the simulated dataset with the observed data, we further generate a set of $x_{\eta}^{obs}$ from $D_{\eta}$ using MDM (\ref{eq: MDM model}) with the given value of $\eta$.  The simulated set of $x_{\eta}^{obs}$ is comparable with the observed dataset $x_{obs}$. Finally, the closeness of  $x_{\eta}^{obs}$ and $x_{obs}$ is evaluated using the KNN distance.  

\begin{figure}[!ht] 
	\begin{center}
		\includegraphics[scale=0.55,angle=-90]{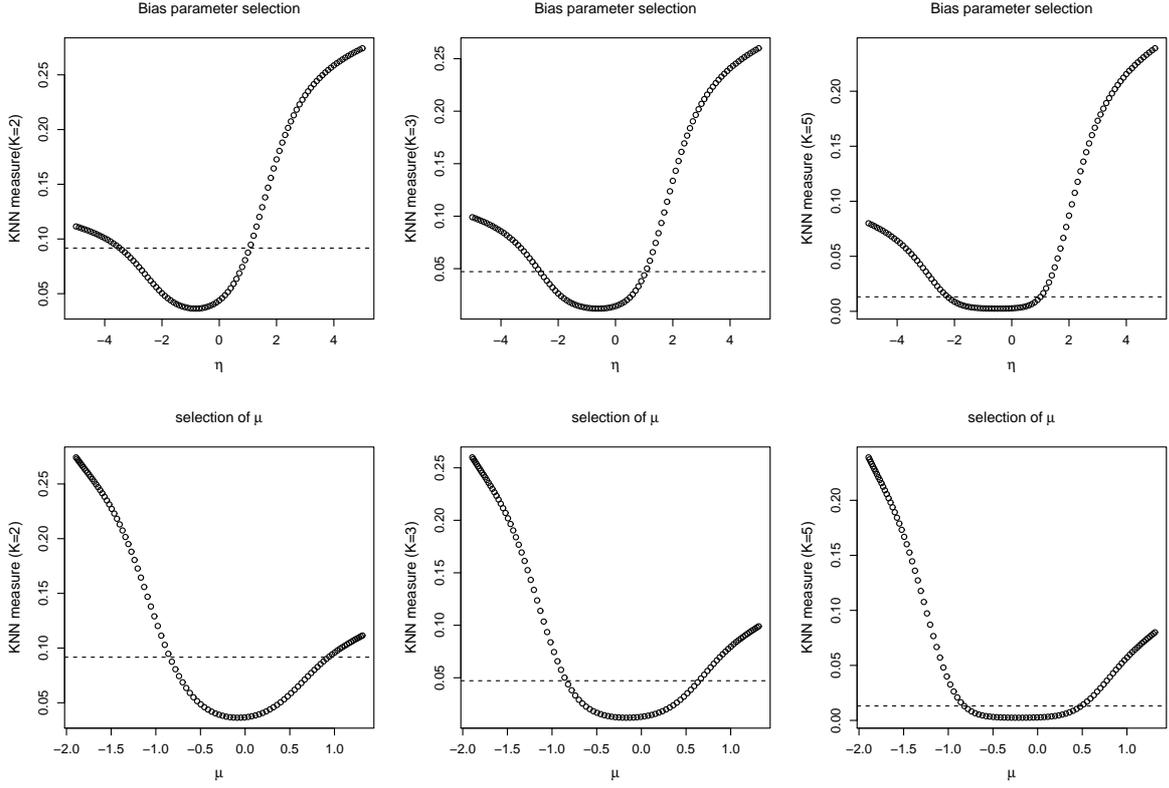}
		\caption{Mean estimation with non-ignorable missing data:  Upper panel -- KNN distances versus different values of $\eta$; Lower panel -- KNN distance versus the corresponding estimate of $\mu$ for the given value of $\eta$. The dashed line indicates the critical values at  5\% significance level.}\label{fig: SA Income}
\end{center}
\end{figure}

Figure \ref{fig: SA Income} shows the results with the MC size of  1000. The KNN distance takes the minimum at $\eta=-0.89$ when $K=2$. The corresponding estimate is $\hat{\mu}=-0.064$ which is quite close to the true value. 
We also used the permutation test to find the plausible set of the sensitivity parameter. The dashed line in Figure \ref{fig: SA Income} indicates the critical value at a significance level of $5\%$; all the ones below that line forms the set of plausible values  resulting in estimates of $\mu$ in the range of (-0.82, 0.91) when $K$=2. We have also repeated the simulation for  other values of $K$. As shown in the same figure, all of them give similar results although the values of the KNN distance is less sensitive to $\eta$ for larger values of  $K$. Further details regarding the selection of $K$  can be found in \citet{HPS08} and \cite{NBBTNM06}. 

Table \ref{table: income psi} presents the simulation study results for 500 replications with different MC sizes and different values of $K$. The value of $\eta$ is selected by the shortest KNN distance with  $\hat{\mu}$ being  the most plausible estimate. If we just use the sample mean of the observed data, the average of $\hat{\mu}$ is about -0.5 which is far away from the true value of 0. Table \ref{table: income psi} shows that the SSA approach gives much better results. This is also supported by the values of coverage probability, which are also reported in Table \ref{table: income psi}.

\begin{table}[!ht] 
\caption{Mean estimation with non-ignorable missing data: simulation study } \label{table: income psi}
\centering 
\begin{tabular}{ccccccccccc} 
\Hline 
& \multicolumn{4}{c}{average of $\hat{\mu}$ (sd)} && \multicolumn{4}{c}{Coverage Probability (\%)} \\ \cline{2-5} \cline{7-10} 
MC size & 10 & 20 & 100 & 1000 & & 10 & 20 & 100 & 1000 \\
\hline

K=2 & -0.065 (0.16) & -0.069(0.16) &-0.064(0.16) &-0.064(0.16) && 80.8 &78.6&89.4&91.6 \\
K=3 & -0.133(0.16) &-0.138(0.16) &-0.158(0.16) &-0.154(0.16) && 73.4 & 72.8 &78.2 &80.4 \\
K=4 & -0.143 (0.17) & -0.152 (0.16) &-0.190(0.16) & -0.191(0.16) && 71.6 & 67.4 &68.6 &68.8 \\
K=5 & -0.169(0.17) &-0.187(0.17) &-0.211(0.16) &-0.204(0.16) && 70.6 & 64.2 &62.0 &66.0 \\ \hline

\end{tabular}
\end{table}

\begin{table}[!ht] 
\caption{Mean estimation with non-ignorable missing data: selection of $\eta$} \label{table: mean esti psi}
\centering 
\begin{tabular}{ccccc} 
\Hline 
& \multicolumn{4}{c}{average of selected $\eta$} \\ \cline{2-5}  
MC size & 10 & 20 & 100 & 1000  \\
\hline

K=2 &-0.85 & -0.87 &-0.88 & -0.89  \\
K=3 & -0.73 & -0.71 &-0.65 &-0.67\\
K=4 & -0.69 &-0.67 &-0.57 &-0.57  \\
K=5 & -0.58 & -0.59 & -0.54 & -0.54  \\ \hline

\end{tabular}
\end{table}

\begin{figure}[!ht] 
\begin{center}
		\includegraphics[scale=0.55, angle=-90]{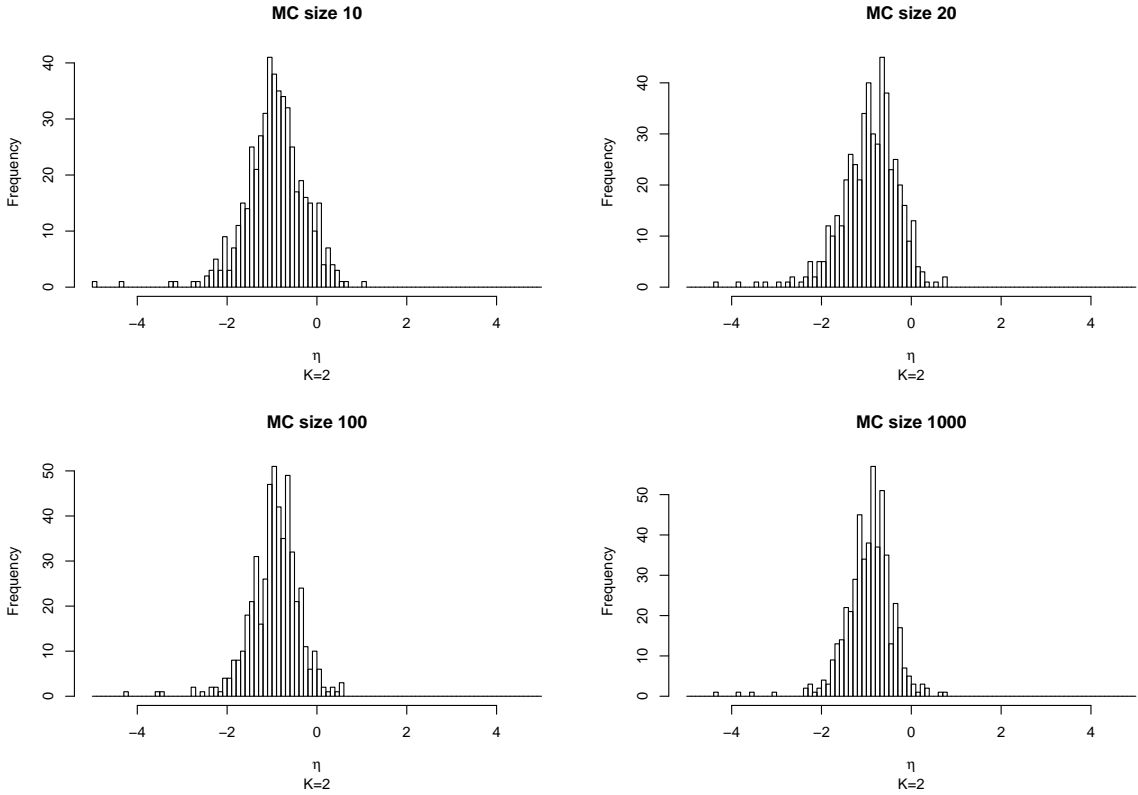}
		\caption{Mean estimation with non-ignorable missing data: histograms of the selected $\eta$ with different MC sizes. }
		\label{fig: SA Incomep}
\end{center}
\end{figure}

The results also  show that the estimates are quite consistent for different values of $K$ even for small number of the MC sizes. Histograms of $\eta$ for a range of MC sizes are provided in Figure \ref{fig: SA Incomep}. As it can be seen, the method with a MC size of 100 or above  when $K=2$ usually gives quite robust results.

\subsubsection*{A.3.2 Regression analysis with missing covariates} \label{sec: semi}

We now consider a regression problem with missing confounders. 
Let $D=(D_{obs}, D_{mis})$, where $D_{obs}$ denotes the data where both the response variable and the covariates are observed; while $D_{mis}$ denotes the data where the response variable is observed but the covariates are only partly observed. We still use $R$ as the missing data indicator and  assume that the MDM  depends on both $D_{obs}$ and  $D_{mis}$. The joint density of data $D$ and the indicator $R$ is expressed by $f(R,D)$, which can be factorized as  $f(R|D)f(D)$. 
We use the following   semiparametric model \citep{KY11} as the selection model for $f(R|D)$: 
\eql{
\mbox{logit}\{P(R=1|D_{obs},D_{mis}) \}=h(D_{obs}) + \eta D_{mis},}{semi} 
where $h(\cdot)$ is a nonparametric function and $\eta$ is a sensitivity parameter. This model takes a nonparametric model on the observed part $D_{obs}$ and a simple linear  form on the missing variable $D_{mis}$. 
Equation \eqref{semi} can be rewritten as
\eql{\pi_D = P(R=1|D) = \expit(h(D_{obs}) + \eta D_{mis}).}{eq: semi mdm}
Using Lemma 1, we have 
\eql{ f(D_{mis}|D_{obs}, R=0) = f(D_{mis}|D_{obs}, R=1) \frac{\exp(-\eta D_{mis})}{E(\exp(-\eta D_{mis})|D_{obs},R=1)}.}{eq: c d selection 1}
The sensitivity parameter $\eta$  determines the amount of departure from the ignorability of the MDM. 
In formula   \eqref{eq: c d selection 1}, we need to compute the conditional distribution of missing data $f(D_{mis}|D_{obs}, R=1)$. A consistent estimate of $f(D_{mis}|D_{obs}, R=1)$ can be obtained parametrically or non-parametrically; an example is shown below. 

To use the SSA approach we first choose a value of $\eta$ from a predetermined set $\Gamma$. For each $\eta$ we simulate  missing data from either \eqref{eq: c d selection} or  \eqref{eq: c d selection 1}. Together, the (simulated) missing  and the observed data form  a dataset which is complete. From here, a subset $D_{\eta}^{obs}$ can be re-sampled using  the MDM specified in \eqref{semi} and the KNN distance between  $D_{\eta}^{obs}$ and the true observed data $D_{obs}$ can finally be calculated.  

We now conduct a simulation study for non-ignorable missing covariate problem based on a real data example. 

The US Federal Highway Administration published fuel consumption data over 50 states and the District of Columbia in 2001; see \citet{Weisberg05} (Chapter 1, page 15). The aim of the research is to understand the effect on fuel consumption ($T$) of several covariates namely, the number of federal-aid highway miles $(X_1)$, the personal income $(X_2)$, the number of licensed drivers  $(X_3)$ and  the state gasoline tax rate $(X_4)$. 

We consider the following linear regression model 
\begin{equation}
t=\theta_0+\sum_{j=1}^4 \theta_j x_j+ \epsilon. 
\label{linreg}
\end{equation}
 By using the complete dataset, estimates for
 $(\theta_0, \theta_1, \theta_2, \theta_3, \theta_4)$ can be obtained and they are  presented in the first row in Table~\ref{table: Fuel sa} . 

We now consider a missing data problem  by letting income ($X_2$) to be partly missing with probability $P(R=0|D) = 1- \expit(1 + (x_1-\bar{x}_1) - 0.5 (x_2-\bar{x}_2)) $, where $R$ is the missing data indicator and $\bar{x}$ is the sample mean. This is a non-ignorable missing data problem.  

To use the SSA approach we resort to the  semi-parametric selection model in \eqref{semi}, i.e. 
\eqn{P(R=1|t,x_1,x_2,x_3,x_4) = \expit(h(t,x_1,x_3,x_4) + \eta x_2).}
Following the discussion around  \eqref{eq: c d selection 1}, we simulate  the missing data from:
\[ f(x_2|t,x_1,x_3,x_4, R=0) = f(x_2|t,x_1,x_3,x_4, R=1) \frac{\exp(-\eta x_2)}{E(\exp(-\eta x_2)|t,x_1,x_3,x_4,R=1)}, \]
where $f(x_2|t,x_1,x_3,x_4, R=1)$ is modelled by the following parametric model:
\[(x_2|t,x_1,x_3,x_4,r=1) \sim N(\gamma_0 +\gamma_1 t + \gamma_2 x_1 + \gamma_3 x_3 +\gamma_4 x_4, \tau^2), \]
where the unknown parameters $(\gamma_0, \gamma_1, \ldots, \gamma_4, \tau^2)$ can be estimated from the observed data and  $X_2$ (personal income) is assumed to be normally distributed.  

The simulated data for missing $x_2$ and the observed data can now be used to form a `complete' dataset $D_{\eta}=(t,x_1, x_2^{*}, x_3, x_4)$  for each given $\eta$. Using $D_\eta$ we can estimate the  parameter $(\hat{\theta}_{\eta}, \hat{\sigma}^2_\eta)$ from the linear regression model \eqref{linreg}. In order to make the SSA method more reliable numerically, we build on the idea of `bootstrapping residuals' introduced in  \citep{ET93}. Instead of  using $D_{\eta}$ directly, we use $D_{\eta}^*=(t^*, x_1, x_2^*, x_3, x_4)$, where $t^*$ is re-sampled conditionally on the following linear regression model with  estimates $(\hat{\theta}_{\eta}, \hat{\sigma}^2)$ and imputed covariates ${x}^*=(x_1,x_2^*,x_3,x_4)^T$:
\[ t^*|x_1,x_2^* , x_3, x_4 \sim N(\hat{\theta}_{\eta} ^T {x^*}, \hat{\sigma}^2_\eta), \] 
i.e. $t^*$ is simulated by adding residuals on the predicted values, where the residuals are sampled from a normal distribution $N(0, \hat{\sigma}^2_\eta)$. And, finally, we then calculate the distance between $D_{\eta}^{obs*}= (t^*, x_1, x_3, x_4)$ and the observed dataset  $D_{obs} = (t, x_1, x_3, x_4)$ by using an average distance a MC size of 100 (this will reduce the  sampling error). 

\begin{figure}[!ht] 
	\begin{center}
		 \subfigure[K=2]{\includegraphics[scale=0.3,angle=-90]{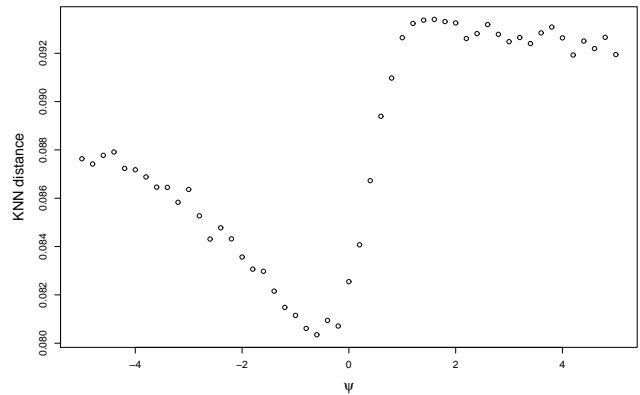}} \\
			\subfigure[K=3]{\includegraphics[scale=0.3,angle=-90]{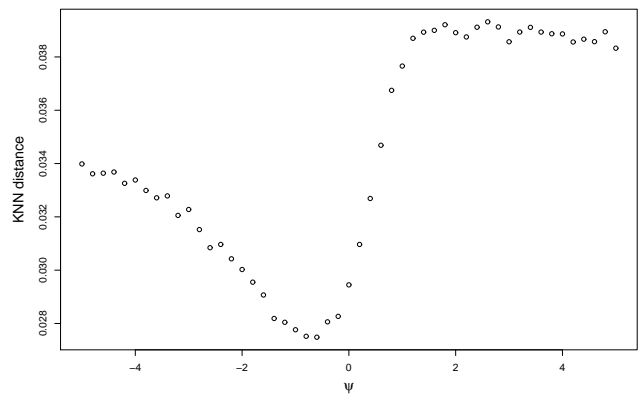}}
		\caption{KNN distance for the fuel consumption data.} 
		\label{fig: Fuel SA}
\end{center}
\end{figure}

In this example  $\eta$ is selected from within the interval $(-5,5)$ at equal-distance of 0.2 unit.  
Figure \ref{fig: Fuel SA}  shows the KNN distances with $K=2$  against the values of $\eta$. As it can be seen, the `most plausible' value or minimum distance is achieved at around $\eta = -0.6$. It is also interesting to note that the corresponding estimates are very close to the ones obtained from the complete data. The ASL using the permutation test has also been calculated but its value is well above the critical value of 0.05 and hence in not showing in the plot.    

 The simulation study results with 100 replications are presented in Table \ref{table: Fuel sa}. The row named as `SSA' shows the average values for the `most plausible estimate'. As seen, all the estimates are very close to the estimates calculated from the complete data except $\hat{\theta}_2$. This is further evidence that the most plausible model obtained via a SSA study needs to be chosen with caution.
  It may provide the `best' results based on the observed data, but it may not be consistent.

 As comparison, we also considered the complete case analysis, using multivariate imputation under MAR (`MICE' Package with Bayesian Linear regression method; see \cite{mice}) and multivariate imputation under MNAR (iteration step for sensitivity analysis, see \cite{SensMice}). Note the latter method modifies the imputation model under a MAR hypothesis to MNAR hypothesis by specifying a supplementary parameter (SupPar). For binary or categorical variables, this parameter is the odds ratio (i.e. the ration between  the odds of the modality interest among subjects with missing values and subjects without missing value). Likewise, for continuous variables, this is the difference in expected values. As a comparison, we have also applied Monte Carlo sensitivity analysis (MCSA) proposed by \citet{Greenland05} 
 and the Bayesian sensitivity analysis (BSA) proposed by \citet{MGL07}. The prior distribution $f(\eta)$ is selected as U(-5,5) for both the MCSA and the BSA methods. As it can be seen in the simulation results with 100 replications presented in Table \ref{table: Fuel sa}, 
overall the SSA method performs more favorably   and robustly than the others. 

\begin{table}[!ht] 
\caption{Simulation study for fuel consumption data} \label{table: Fuel sa}
\centering 
\begin{tabular}{l c c c c c} 
\Hline 
  & $\hat{\theta}_0$ (se) & $\hat{\theta}_1$ &  $\hat{\theta}_2$ & $\hat{\theta}_3$ & $\hat{\theta}_4$  \\
\hline 
Complete data &154.19(9.08) & 18.55(6.47) & -6.14(2.19) & 0.47(0.13) & -4.23(2.03)\\
Complete case & 155.06(14.44) & 16.20(12.41) & -7.32(3.37) & 0.24(0.18) & -7.29(2.24) \\  \\
Multiple Imputation: &&&&& \\
	MAR 						& 130.15(13.11) &30.68(8.5)   &-9.86(4.13)  &0.50(0.15)  &-5.17(2.17) \\  
	MNAR (SupPar=1) & 137.70(12.03) & 24.24(8.95) &-9.74(3.91)  & 0.43(0.14) & -5.42(2.37) \\
	MNAR (SupPar=2) & 141.26(11.27) & 22.84(8.87) & -9.68(3.65) & 0.42(0.14) &-5.33(2.34) \\
	MNAR (SupPar=5) & 152.11(9.69)  &19.45(8.27)  &-8.19(2.88)  & 0.42(0.14) & -5.01(2.24) \\
	MNAR (SupPar=8) & 159.49(9.36)  & 17.69(7.73) & -6.34(2.24) & 0.43(0.14) &-4.75(2.17) \\ \\
Sensitivity Analysis: &&&&& \\	
		SSA (K=2)			& 152.30(8.29)  &18.24(5.59)  &-3.99(2.18)  & 0.49(0.12) &-4.39(1.85) \\ 
		SSA (K=3)     & 154.12(8.22)  &17.01(5.57)  &-4.05(2.08)  & 0.49(0.11) &-4.38(1.83) \\
		MCSA 					&153.62(13.22)  &22.87(6.94)  &-3.44(1.87)  &0.49(0.13) &-4.29(2.07) \\
	  BSA 					&131.51(10.58)  &31.77(6.49)  &-10.17(2.82) &0.45(0.12) &-4.75(2.00) \\ 
\hline 
\end{tabular}
\end{table}

\end{document}